\documentclass[prl,showpacs,amsmath,amssymb,twocolumn]{revtex4}

\usepackage[dvips]{epsfig}
\usepackage[dvips]{graphicx}
\usepackage{dcolumn}% Align table columns on decimal point
\usepackage{bm}% bold math

\graphicspath{{converted_graphics/}}
\begin{document}

\title{Efficient multiqubit entanglement via a spin-bus} 

\author{Mark Friesen$^1$}
\email{friesen@cae.wisc.edu}
\author{Asoka Biswas$^{2}$}
\author{Xuedong Hu$^{3}$}
\author{Daniel Lidar$^{2}$}
\affiliation{$^1$Department of Physics, 
University of Wisconsin-Madison, Wisconsin 53706, USA} 
\affiliation{$^{2}$Departments of Chemistry, Electrical Engineering and Physics, 
University of Southern California, Los Angeles, California 90089, USA}
\affiliation{$^{3}$Department of Physics, University at Buffalo, 
State University of New York, Buffalo, New York 14260-1500, USA}

\begin{abstract}
We propose an experimentally feasible architecture with controllable long-range
couplings built up from local exchange interactions.  The scheme
consists of a spin-bus, with strong, always-on interactions, coupled dynamically to external 
qubits of the Loss and DiVincenzo type.
Long-range correlations are enabled by a spectral gap occurring in a finite-size chain.
The bus can also form a hub for multiqubit entangling operations.  We show how  
multiqubit gates may be used to efficiently generate $W$-states (an
important entanglement resource).  The spin-bus therefore provides a 
route for scalable solid-state quantum computation,
using currently available experimental resources.
\end{abstract}


\maketitle

Spin qubits in quantum dots are considered leading candidates for quantum computation
because of their long decoherence times and their affinity to scalable gating 
techniques \cite{loss98,elzerman05}.  A prominent form of interaction between
the spins is the exchange coupling, which has long been
considered crucial for quantum computing in quantum dots 
\cite{loss98} and other spin-based qubits \cite{kane98}.
By utilizing coded qubits, the exchange 
coupling can even accomodate \textit{single} qubit rotations 
\cite{bacon00,kempe01}.  Recent experiments lends support to
the Loss \& DiVincenzo proposal by demonstrating electrical 
control of the exchange coupling in spin qubits \cite{petta05}.  
However, the range of the interaction is only tens of 
nanometers, leading to scaling and architectural constraints \cite{svore05}. 
In bulk systems, an effective interaction, known as 
Ruderman-Kittel-Kasuya-Yosida or RKKY \cite{kittelbook}, arises due to the exchange
coupling between localized moments and intermediary particles, such as electrons in
a commonly shared electron gas \cite{craig04}
or virtual excitons \cite{Piermarocchi}.  The resulting, effective
interactions are long-range.    
It is presently unknown whether a spin chain could play a 
similar intermediary role for electron spins in quantum dots, 
thereby enabling a more scalable quantum computing architecture.
Here, we show how to engineer such long-range interactions between remote qubits 
connected to a spin-bus, and we identify appropriate bus gate operations. 
We also show how critical quantum resources, like an entangled
many-body $W$-state, can be generated efficiently by means of a spin-bus.
We find that long-range, many-body interactions
can be achieved using established, electrically controlled gating methods.

Architectures that use an intermediary bus to facilitate 
long-range interactions between remote qubits
have been studied in various qubit schemes, with the Cirac-Zoller proposal for 
trapped ions as a preeminent example \cite{CiracZoller}.  In semiconductors, there 
have been bus proposals to transduce spin information into photon modes  
in resonant cavities \cite{imamoglu99}
or transmission lines \cite{childress04}.  
Spin chains have been proposed as a medium for long-range correlations 
\cite{venuti} and dynamical modes \cite{yung06}.  
However, the adiabatic spin-bus remains unexplored in the context of 
quantum computation.  

A spin-bus naturally combines long-range interactions with the
connectivity needed for computations.  Several versions of the bus are shown in
Fig.~1.  The bus itself is formed of individual electron spins in a chain with strong, 
static exchange couplings.  External spin qubits are coupled dynamically to the internal 
nodes of the bus by means of electrical gates.  
We anticipate that a dedicated register may be optimized for the special needs 
of the bus.  However, the main physical requirements are no different from ordinary quantum
dots in the Loss \& DiVincenzo qubit scheme.  

Spin interactions within the bus are described by the Hamiltonian
$H_b=J_b\sum_{i=1}^{N-1}
\mathbf{s}^b_i\cdot\mathbf{s}^b_{i+1}$.  Here, the bus spin operators $\mathbf{s}^b_i$
act on the constituent spins, and we assume the bus size $N$ is strictly odd.  
(An $N=2$ bus is considered in \cite{kane98}.)
We take the internal bus couplings
$J_b$ to be uniform, although this is not essential for bus operation.
When $N$ is odd, the bus spectrum exhibits a spin-$1/2$ doublet ground state, 
separated from the excited states by a spectral gap \cite{meier03,li05}
\begin{equation}
\Delta_b \simeq J_b\pi^2/2N ,
\end{equation}  
due to finite system size \cite{dagotto96}.
This ground state manifold, spanned by $\{ |0\rangle_b,|1\rangle_b \}$, 
forms the working space of 
the bus, as illustrated in Figs.~2(a) and 2(b).  If the ``adiabatic temperature" 
$T_{{\rm gate}}\sim 1/\tau_{{\rm gate}}$ (for gate period $\tau_{\rm gate}$) 
and the physical temperature are both smaller than the minimum gap,
$\Delta_{\rm min}\simeq \Delta_b$, then the bus will remain in its 
working manifold, once initialized.  The coupling between the $i$th qubit and the $i$th bus 
spin is given by $H_i=J_i(t)\,\mathbf{s}^q_i\cdot\mathbf{s}^b_i$.  Restricting the bus to its
ground state manifold, we obtain an effective qubit-bus Hamiltonian: 
\begin{eqnarray} 
&& \hspace{.05in} H^*_i=J^*_i(t)\,\mathbf{s}^q_i\cdot\mathbf{S}, \\
&& J^*_i= 2\,J_i\,{}_b\langle 1|s_{iz}^b|1\rangle_b ,
\end{eqnarray} 
where the spin operator $\mathbf{S}$ acts on the spin-$1/2$ bus manifold.  
Numerically, we find that 
$J^*_i\simeq \pm J_i/\sqrt{N}$, where the $+(-)$ sign holds for odd (even) bus nodes.
Thus, the qubit-bus coupling strength $J_i$ and the size of the bus determine its 
operating speed.  We note that the effective couplings $J_i^*$ alternate
between ferromagnetic and antiferromagnetic, reminiscent of the RKKY interaction, while
the effective coupling strength decays as a power-law, also similar to RKKY.  Below, we shall 
assume uniform qubit-bus couplings ($J_i=J_q$, for all $i$) and consider only 
the antiferromagnetically coupled nodes.

The simplest operating mode of the bus is the serial mode, in which the bus acts as
a qubit proxy.  The Heisenberg interaction $H^*_i$ generates
a SWAP gate between the qubit and the bus \cite{loss98}, after a gate time 
$\tau_{{\rm SWAP}}\simeq \pi\sqrt{N}/J_q$.  
The serial gate protocol proceeds as follows:  (i) SWAP qubit onto bus, (ii) perform
root-SWAP gate between bus and target qubits, (iii) SWAP bus back onto original 
qubit.  The ideal final state corresponds 
to a root-SWAP between the qubits, leaving the bus in its initial state.
Note that the initial state of the bus is irrelevant for
serial operations.  However, it must be initialized into its working manifold.  This can be 
accomplished quickly by thermalization, when the gap $\Delta_b$ is much larger than 
the temperature.  

\begin{figure}[tbp] % float placement: (h)ere, page (t)op, page (b)ottom, other (p)age
  \centering
  % file name: C:/PCTeXv4/LaTeX/busgeom.eps
  \includegraphics[bb=69 170 551 776,width=2.5in,keepaspectratio]{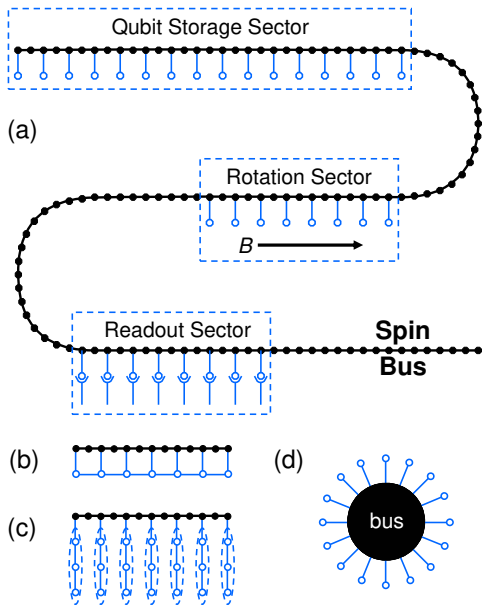}
  \caption{(Color online)
Spin-bus architectures.  
(a) The spin-bus is a chain of electronic spins 
(closed circles) with strong, static couplings (heavy lines).  
External qubits (open circles) can be coupled to the bus  
at any node (light lines).  Effective long-range interactions 
allow for communication between sectors dedicated to
rotation, read-out or memory, which may benefit from isolation.  
(b) Additional local couplings enable parallel 
interactions, in addition to bus-mediated interactions.
(c) Coded qubits or larger clusters.
(d) Within the ground state manifold, the bus acts as a simple 
spin-$1/2$ qubit, except for its plurality of qubit couplings.}
  \label{fig:busgeom}
\end{figure}

The scaling properties of the serial operating mode are determined by the bare coupling
constants $J_b$ and $J_q$, and the bus size $N$.  Using the relations given above, the
adiabaticity criterion, $2\pi < \tau_{{\rm SWAP}}\Delta_b$, can be rewritten as
$J_b/J_q > 4\sqrt{N}/\pi^2$.  Scaling up to large $N$ therefore depends on arranging for 
a large 
ratio between the coupling constants.  Because of the exponential dependence of the exchange
coupling on the quantum dot separation and the barrier height, one can easily imagine a 
coupling constant ratio of order $J_b/J_q > 100$.  This suggests a bound of
$N< 60,000$ for the bus size, corresponding to a gap of $\Delta_b=1$~mK 
when $J_b=1$~meV.  Alternatively, the gap increases to 100~mK when $J_b=2$~meV, 
for a bus of size $N<1,200$.

\begin{figure}[t] % float placement: (h)ere, page (t)op, page (b)ottom, other (p)age
  \centering
  % file name: C:/PCTeXv4/LaTeX/spectrumfig.eps
  \includegraphics[bb=32 214 577 781,width=2.7in,height=2.7in,keepaspectratio]{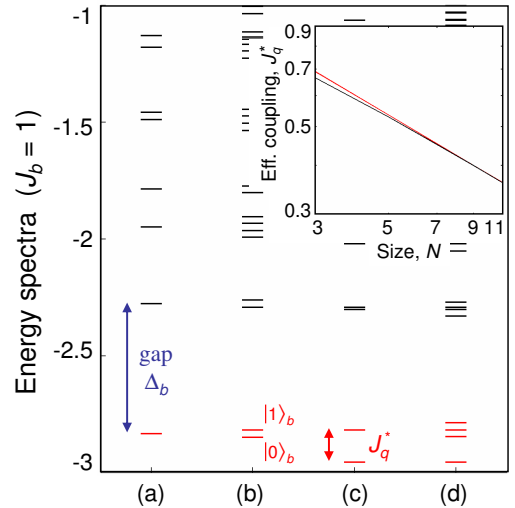}
  \caption{(Color online)
Energy spectra for an $N=7$ bus.  
The ground state doublet (the bus manifold) lies below the gap (red lines).  Excited states
lie above the gap (black lines).
(a) No coupled qubits, $B=0$:  bus manifold is doubly degenerate.
(b) No coupled qubits, $B=0.03$:  the doublet splits, defining the
working states of the bus.
(c) One coupled qubit, $B=0$:  bus and qubit states hybridize  
to form a singlet and a triplet, split by $J_q^*$.  
(d) One coupled qubit, $B=0.03$:  the triplet states split.
In (a)-(d), we use $J_b=1$ and $J_q=0.3$, with the dimensionless 
Zeeman coupling $H_B=\sum_iB s_{iz}$, summed over all spins.
Inset:  Log-log plot of the effective coupling $J_i^*$ from Eq.~(3), 
averaged over the antiferromagnetic bus nodes (lower, black curve),
with bare couplings set to $J_i=1$.  Upper (red) curve shows apparent
asymptotic behavior, $J_q^*\simeq 1.198 N^{-1/2}$.}
  \label{fig:spectrumfig}
\end{figure}

To compare the scaling properties of the serial bus gate to a conventional
linear qubit array, we consider a 
SWAP protocol between two qubits on opposite ends of an $N$-qubit chain.   
For the conventional array, this involves a series of $(2N-3)$ SWAP gates.  Since some
gates of duration $\pi/J_q$ may be performed in parallel, the total gate time is roughly
$N\pi/J_q$.  The corresponding spin-bus protocol involves just three SWAPs, 
with a total gate time
of $3\pi\sqrt{2N-1}/J_q$, where we have assumed a $(2N-1)$-qubit array,
with $N$ antiferromagnetically-coupled qubits and $(N-1)$ unused 
ferromagnetically-coupled qubits.
Thus, for a serial SWAP gate, the bus provides a quadratic speedup.  
A modest amount of parallel (qubit-qubit) connectivity [Fig.~1(b)] also 
enables local gates, and parallel gate operations.

We can compare the propagation of errors in SWAP protocols by 
introducing small random errors of magnitude $\delta$ into the couplings:  
$J_q\rightarrow \tilde{J}_q =J_q(1\pm \delta)$.  If $U_{\rm SWAP}(J_q)$ represents
a perfect multiqubit SWAP gate, then the operator norm
$\epsilon=||U_{\rm SWAP}(\tilde{J}_q)-U_{\rm SWAP}(J_q)||$ describes the compounded
error.  We have performed numerical simulations of $\tilde{J}_q$ errors in a conventional 
linear qubit array.  By averaging over random error realizations,
we observe that the resulting errors add up as a type of random walk, 
with $\epsilon \propto \delta(2N-3)^{0.68}$.
An equivalent analysis for the spin-bus shows that
$\epsilon$ does not depend on $N$, leading to a scaling improvement of particular significance
for quantum error correction \cite{svore05}.

In the serial mode, the spin-bus functions as a simple conduit for quantum
information, leading to a new scaling law for long-range gating.
However, the full potential of the spin-bus is achieved through
simultaneous, multiqubit couplings, which enable quantum-parallel-processing.  
As an example, we now show how to efficiently generate  
$W$-states of $n$ qubits \cite{duer00}, defined by 
$|W_n\rangle=(|00\dots 001\rangle+|00\dots 010\rangle+|00\dots 100\rangle+\cdots+
|10\dots 000\rangle)/\sqrt{n}$.
Such highly entangled states form a critical resource for quantum computation
because of their robustness to particle loss \cite{haeffner05}
and their relative immunity to dephasing \cite{roos04}.

We first show how to construct multiqubit bus gates.
Simultaneous multiqubit couplings to the bus are described by the Hamiltonian
$H_n=J_q^*\sum_{i=1}^{n}\mathbf{s}_i^q\cdot\mathbf{S}$.  Here, the bus behaves as 
an ordinary qubit when it is restricted to its working manifold,
except for its plurality of couplings [Fig.~1(d)].  
Here we assume the effective coupling constants $J_q^*$ are identical for all 
qubits, although lifting this restriction enables a richer set of multiqubit gates.

The unitary evolution operator
for the qubit-bus system is given by $U(t)=e^{-iH_nt}$.  Although $U(t)$
possesses off-diagonal terms that entangle the qubits with the bus, these terms 
should vanish for a true bus gate.  We therefore seek  
special evolution periods $t=\tau$ for which bus decoupling occurs. 
The task of computing $U(t)$ is simplified in the angular momentum basis
$\{|0\rangle,|1\rangle\} \otimes \{|j,\lambda,m\rangle\}$, where the states on the left
describe the bus manifold, and the qubit states on the right are classified by their
angular momentum quantum numbers, $j$ and $m$, and the degeneracy
label $\lambda$ \cite{kempe01}. 
(For example, in the case of four qubits, there are two orthogonal
$j=m=0$ singlet states; whence $\lambda=0,1$.)  
In the angular momentum basis, $U(t)$ is block diagonal with blocks of size 
$1\times 1$ and $2\times 2$.  The latter correspond to pairs of states given by
$\{ |0\rangle |j,\lambda,m+1\rangle ,|1\rangle |j,\lambda,m\rangle \}$.  We find
there is a time $\tau$ for which all the $2\times 2$ blocks are simultaneously
diagonal, given by
$\tau=4\pi/J_q^*$ when $n$ is even, and $\tau=2\pi/J_q^*$ when 
$n$ is odd.  (The case $n=2$ is anomalous, with $\tau=4\pi/3J_q^*$.)  The resulting 
diagonal bus gates are
\begin{eqnarray}
&& \langle 0;j,\lambda,m|U(\tau)|0;j,\lambda,m\rangle= 
\\ \nonumber  && \hspace{.4in}
\langle 1;j,\lambda,m|U(\tau)|1;j,\lambda,m\rangle=e^{-ijJ_q^*\tau/2} .
\end{eqnarray}   
When $n$ is even, we find that $U(\tau)=\mathbf{1}$ (except when $n=2$).  
However, the case of odd $n$ produces nontrivial bus gates.  
Since the multiplicity of each diagonal 
element in $U(\tau)$ is even, we may reorder the basis such that
$U(\tau)=\textrm{diag}(U_n,U_n)=\mathbf{1}_b \otimes U_n$.
This is a remarkable result:  the action of the
$U(\tau)$ gate is to return the bus to its original state, while implementing
a non-trivial transformation $U_n$ on the qubits.  On the other hand, imperfect
gate timing may produce unwanted entanglements between the qubits and the bus.
For small timing errors of the form $J_q^* \tilde{\tau} = 2\pi (1+\delta )$, we compute
the operator error norm $\epsilon=||U(\tilde{\tau})-U(\tau)||=(\pi/2)(n+2)\delta$,
and the state fidelity, 
$1-f=1-{\rm Tr}\left\{ \left[U(\tau)|\Psi\rangle\langle\Psi|U^\dagger(\tau)\right]
\left[U(\tilde{\tau})|\Psi\rangle\langle\Psi|U^\dagger(\tilde{\tau})\right] \right\}
<(\pi^2/2)(n+1)^2\delta^2$.

The temporal scaling properties of the bus gate derive from the fact that $\tau$ is
independent of $n$.  Thus, in terms of time resources, multiqubit bus gates cost  
the same as few-qubit bus gates.  In contrast, many time steps are needed 
when building conventional, multiqubit operations out of local gates, 
especially when quantum error correction is taken into account.
The spatial scaling properties of the bus gates are determined by diagonalizing 
the Hamiltonian $H_n$.  The resulting spectrum has width 
$\delta E = (1+n)J_q^*/2$, which grows linearly with the number of coupled qubits.  
For adiabatic operation, this spectrum should lie entirely
below the gap:  $\delta E < \Delta_b$.  We consider $U_n$ acting on $n$ 
antiferromagnetically-coupled qubits, corresponding to a minimum bus size of $(2n-1)$.  The
resulting bound on the gate size is $n< (\pi^2J_b/J_q\sqrt{2})^{2/3}\simeq 79$, 
where we have used our previous, conservative estimate of $J_b/J_q\simeq 100$.  

We now develop protocols for generating $W_n$ states, using the multiqubit 
gates $U_n$.  The probablistic procedures require the measurement of certain
``sacrificial" ($s$) qubits.  Following a successful measurement
outcome, the remaining ``data" ($d$) qubit register is found in the desired 
$W_n$ state.  A whole family of protocols can be derived, involving a variable number 
of sacrificial qubits.  

Two optimal protocols stand out.  In the first case, just
one sacrificial qubit is used.  Consider the specific case of two data qubits.
We find that the bus gate $U_3$, operating on the
initial state $|00\rangle_d|1\rangle_s$, gives 
$(\sqrt{8}/3)|W_2\rangle_d |0\rangle_s -(1/3)|00\rangle_d |1\rangle_s$.  If 
measurement of the sacrificial qubit gives
$|0\rangle_s$, then the data register will be found
in the state $|W_2\rangle_d$.  The protocol can be extended to any system size, 
obtaining $|W_n\rangle$ with probability $p=4n/(n+1)^2$.  
(See \cite{plenio05} for a related quantum oscillator protocol.)  

In the second
case, we use $(n-1)$ sacrificial qubits in an $n$-qubit data register.
The resulting success rate is much higher, but at the cost of 
some qubits.
The protocol can be expressed in the computational basis as follows:
\begin{eqnarray} &&
U_{2n-1}|1\rangle_d^{\otimes n}|0\rangle_s^{\otimes (n-1)}
\stackrel{M_s}{\longrightarrow}
|W_n\rangle_d|1\rangle_s^{\otimes (n-1)}, 
\label{eq:protocol} \\ \nonumber && \hspace{1in}
\{p=n[(2n-2)!!/(2n-1)!!]^2\},
\end{eqnarray}
where $M_s$ signifies the measurement of the sacrificial qubits, and $p$ is the 
probability of success.
In words, the qubits are initialized to a simple product state of 1's in the data
register and 0's in the sacrificial register.  
An entangling $U_{2n-1}$ gate is performed on the combined
register, followed by a parallel measurement of all the qubits in the sacrificial register.  With high probability $p$, the 
sacrificial register will be found in the state with all 1's, with the data register in the 
desired state $|W_n\rangle_d$.  To compute $p$, we express the initial state in the
angular momentum basis.  After applying $U_{2n-1}$, we return to the computational
basis using Clebsch-Gordon techniques.  
We see that $p>\pi/4\simeq 0.79$ for all $n$, so on average,
only 1-2 iterations are needed for a successful outcome.  In comparison,
a deterministic protocol for generating $|W_n\rangle$ via local interactions involves
a series of $n$ exchange gates, and an attendant overhead for 
quantum error correction.

Finally, we consider the decoherence properties of a spin-bus.  The combination of
a large bus size and its always-on couplings leads to
decoherence mechanisms that differ from single-spin qubits.
For example, fluctuations in the inter-bus coupling constant $J_b$ do not cause dephasing 
of the bus state, in contrast with single-spin qubits \cite{chargefluctuations}.
Instead they lead to relatively weak fluctuations of the gap, $\Delta_b$.  
The main dephasing mechanisms for the 
spin-bus are fluctuations of the qubit-bus couplings $J_q$, and the locally 
and temporally varying magnetic fields at the bus nodes, arising from nuclear 
spins \cite{meier03,Merkulov}.  In general, we expect better decoherence properties 
from a spin-bus than an equivalent array of single-spin qubits.  For the bus, 
decoherence rates will scale as $\sqrt{N}$ or $N$ when the nuclear dynamics result in
$1/f$ or gaussian noise, respectively.  
However, the true physical dynamics is not known at present.

We have shown that the exchange coupling possesses a significant untapped potential for quantum
dot quantum computing in the form of long-range interactions via a spin-bus.
The main scaling properties depend on the ratio between the raw qubit and bus coupling 
constants, $J_q/J_b$, and may allow for bus sizes greater than $10^3$.  
But while long-range couplings between spin qubits are beneficial, 
the true power of the spin-bus originates from quantum many-body physics.  
To utilize this potential, we have shown how 
to generate entangling gates, $U_n$, and multiqubit $W_n$ states.  For both serial
and multiqubit operations, the spin-bus provides a new scaling power law for
spin-based quantum computing. 

Work supported by NSF Grant Nos.\ CCF-0523675 and CCF-0523680, and ARO/NSA Contract No.\
W911NF-04-1-0389.

\end{document}